\documentclass{iau}

\usepackage{amsmath}
\usepackage{graphicx}
\usepackage{multirow}
\usepackage{wrapfig}

\newcommand{\Fig}[1]{Figure~\ref{#1}}
\newcommand{\Tab}[1]{Table~\ref{#1}}

\begin{document}

\lefttitle{Cambridge Author}
\righttitle{Proceedings of the International Astronomical Union: \LaTeX\ Guidelines for~authors}

\jnlPage{1}{7}
\jnlDoiYr{2021}
\doival{10.1017/xxxxx}

\aopheadtitle{Proceedings IAU Symposium}
\editors{A. V. Getling \&  L. L. Kitchatinov, eds.}

\title{Exploring the predictability of the solar cycle from the polar field rise rate: Results from observations and simulations}

\author{Akash Biswas}
\affiliation{Department of Physics, Indian Institute of Technology (Banaras Hindu University), Varanasi 221005, India}

\begin{abstract}
The inherent stochastic and nonlinear nature of the solar dynamo makes the strength of the solar cycles vary in a wide range, making it difficult to predict the strength of an upcoming solar cycle. Recently, our work has shown that by using the observed correlation of the polar field rise rate with the peak of polar field at cycle minimum and amplitude of following cycle, an early prediction can be made. In a follow-up study, we perform SFT simulations to explore the robustness of this correlation against variation of meridional flow speed, and against stochastic fluctuations of BMR tilt properties that give rise to anti-Joy and anti-Hale type anomalous BMRs. The results suggest that the observed correlation is a robust feature of the solar cycle and can be utilized for a reliable prediction of peak strength of a cycle at least 2 to 3 years earlier than the minimum.
\end{abstract}


\maketitle

\section{Introduction}
The Sun's magnetic fields dominates and determines the space weather conditions of the heliosphere. The interaction of the magnetic fields embedded in the coronal mass ejections from the Sun with the Earth's magnetosphere gives rise to various phenomena in near-Earth space that can have serious consequences on our lives. The geomagnetic storms that occur through these interactions can damage our space and ground-based communication systems, impact the power grids and can pose a challenge to our space exploration capabilities \citep{Gopal22}. Hence, the study of space weather and an adequate understanding of the evolution of solar magnetic fields is crucial for taking necessary measures to safeguard assets from the impact of these adverse conditions. 

In this context, the forecast of space weather is incomplete without the capability of reliably predicting the strength of an upcoming solar cycle well in advance. The solar cycles are driven by the action of the dynamo operations that facilitate the cyclic conversion of strength from the poloidal component of the magnetic field to the toroidal field and back \citep{Kar14a, Cha20, Hat15, Karak23}. The conversion of the toroidal field strength into the poloidal field exhibits some nonlinear mechanisms \citep{BKC22, Jha20, J20, Kar20} and stochastic fluctuations due to the randomness in the properties of the Bipolar Magnetic Regions (BMRs) \citep{BKUW23, KM17, Jha20, SJKB23}. Due to the combined impact of these processes, the strength and other features of the solar cycles vary widely from one cycle to the other, which makes the job of predicting an upcoming solar cycle incredibly difficult \citep{Petrovay20, kumar21b, Bhowmik23}. 

Among the various methodologies used for the prediction of solar cycle strength, the polar precursor method has gained consensus in the recent times owing to its success in providing successful prediction of Cycle 24 and convergence in the predicted strength of Cycle 25 \citep{CCJ07, Bhowmik+Nandy, Nandy21, UH18, Jiang23}. However, one major drawback of this method is that, the polar field strength at around the solar cycle minimum is used as an input for a reliable prediction of the upcoming solar cycle, this significantly limits the time window of the forecast. 

\cite{Pawan21} reported that the direct correlation of the polar precursors with the strength of the following cycle can be used for cycle prediction after 4 to 5 years of the polar field reversal. Recently, \cite{KBK22} analysed the observed polar field data and showed that, there exists a strong correlation between the rise rate of the polar field and the peak of polar field at the cycle minimum as well as the strength of the upcoming cycle. As the sample size of the observed data is limited to only last three solar cycles, this correlation is not yet statistically robust. However, if this correlation stays high for a larger set of cycles, it can enable us to make reliable prediction of the upcoming solar cycle much ahead of the cycle minimum.

In this work, we explore the reliability of the correlation of the polar field rise rate with the peak of the polar field and the peak strength of the upcoming cycle under the impact of the stochastic properties of the BMRs, especially in the presence of `rogue' (or anti-Hale and anti-Joy) BMRs  through the Surface Flux Transport (SFT) simulations.

\section{Surface Flux Transport Model}
The primary objective of the SFT model is to simulate the build-up of the Sun's polar field through the Babcock-Leighton mechanism \citep{Ba61, Leighton69} for the evolution of radial component of the surface magnetic field from the decaying BMRs under the action of meridional circulation, differential rotation, and horizontal diffusion by numerically solving the following equation:
\begin{eqnarray}
\frac{\partial {B_r}}{\partial t} = - \Omega(\lambda)\frac{\partial B_r}{\partial \phi} - \frac{1}{R_\odot \cos\lambda}\frac{\partial {}}{\partial \lambda} \left[ v(\lambda)B_r \cos\lambda \right] + 
\nonumber\\
\eta_H  \left[ \frac{1}{R_\odot ^2 \cos\lambda}\frac{\partial {}}{\partial \lambda}\left( \cos\lambda \frac{\partial {B_r}}{\partial \lambda} \right) + \frac{1}{R_\odot ^2 \cos^2\lambda}  \frac{\partial^2 {B_r}}{\partial \phi^2}  \right] + 
\nonumber\\
D(\eta_r) + S(\lambda, \phi, t).
\label{eq:ind2}
\end{eqnarray} 

For this study, we use the SFT code that has been used in multiple previous studies like \cite{Bau04, CJSS10} to name a few. We use synthetic spatio-temporal profile of the solar cycles and the BMRs as prescribed by \cite{Jiang18} and \citet{HWR94}. For the calculation of the toroidal flux to be produced by the action of the differential rotation on the polar field developed in the SFT simulations, we use the following equation as prescribed by \cite{CS15}:

\begin{equation}
    \frac{\rm d\Phi^N_{tor}}{\mathrm {d} t~~~~} = \int_{0}^{1}(\Omega - \Omega_{eq})B_rR^2_\odot \mathrm{d}(\cos\theta) - \frac{\Phi^N_{tor}}{\tau}
    \label{tor}
\end{equation}
More details about the SFT model parameters, properties of the BMRs, and the methodology of analysis of the simulated data can be found in \cite{BKK23}.

\section{Results and Discussion}
In the build-up of the poloidal field of the Sun through the Babcock-Leighton mechanism, the meridional circulation plays a major role in transporting the remnant radial surface fields from the decaying BMRs towards the polar regions. Hence any variation in the meridional circulation can change the rate of the production of the polar field and other features of the solar cycle \citep{Kar10, KC11, KC12}. On the other hand, the properties of the BMRs, especially their tilts are a major factor in determining their contribution towards the polar field build-up. Hence, the observed large scatter in the tilts ($\gamma$) of the BMRs around the famous Joy's law \citep{Hale19} can be a major factor in producing fluctuation in polar field strength. In extreme cases, the tilt of some BMRs is opposite of that of the convention for that particular cycle, these BMRs with unusual properties are named as the `rogue' BMRs which are mainly of two types depending on the value of their tilt, the anti-Joy ($-90^\circ<\gamma<0^\circ$) and anti-Hale BMRs ($-180^\circ<\gamma<-90^\circ$) \citep{McClintock+Norton+Li14, munoz21}. Earlier studies have shown that these `rogue' BMRs can severely impact the build-up of the polar field in a negative sense and can create a challenge for the predictability of solar cycle strength \citep{Nagy17,GBKKK23, Pal23}. 

\begin{figure}
\centering
\includegraphics[scale=0.28]{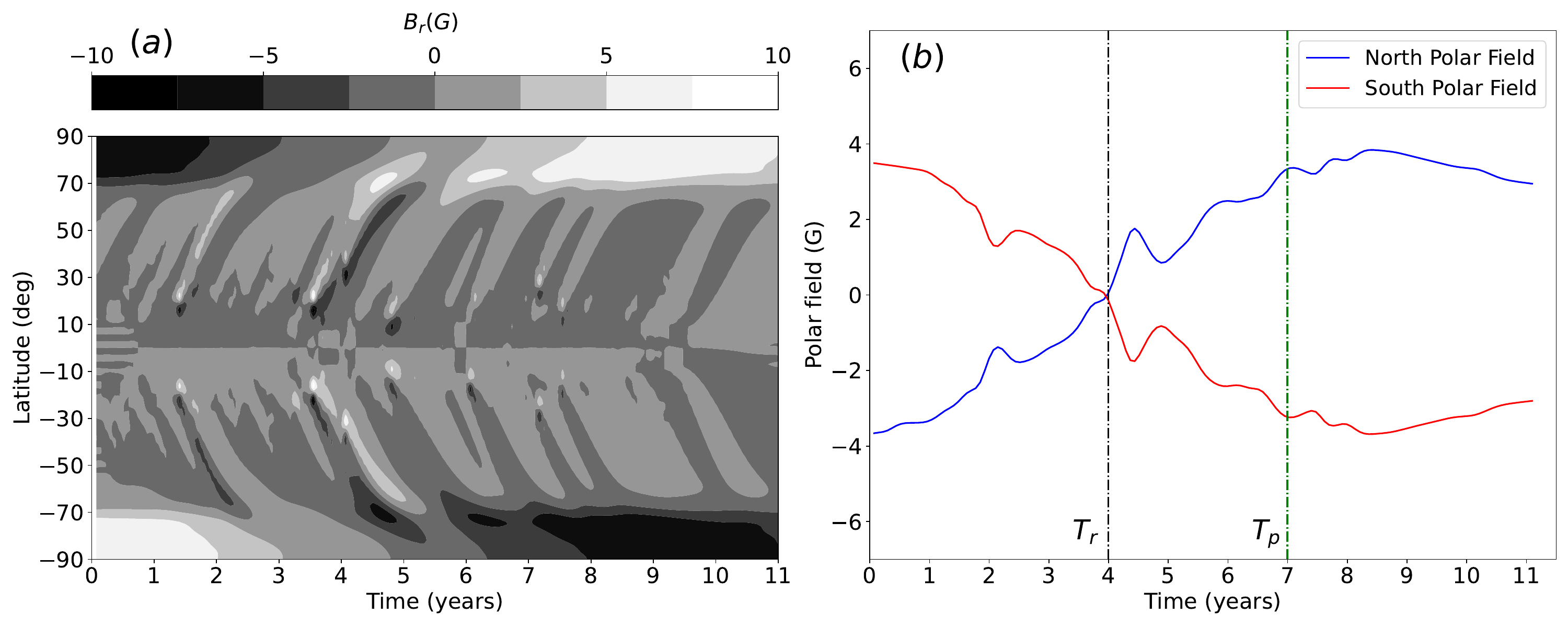}
\caption{ (a) The evolution of the surface radial field under the impact of the surface flows and horizontal diffusion. (b) Evolution of the polar field under the impact of the stochastic properties of the BMRs. The black and green lines represent the time of polar field reversal ($T_r$) and the time of prediction ($T_p$) of the upcoming cycle's strength respectively.
}
\label{fig:pol}
\end{figure}

\begin{figure}
\centering
\includegraphics[scale=0.41]{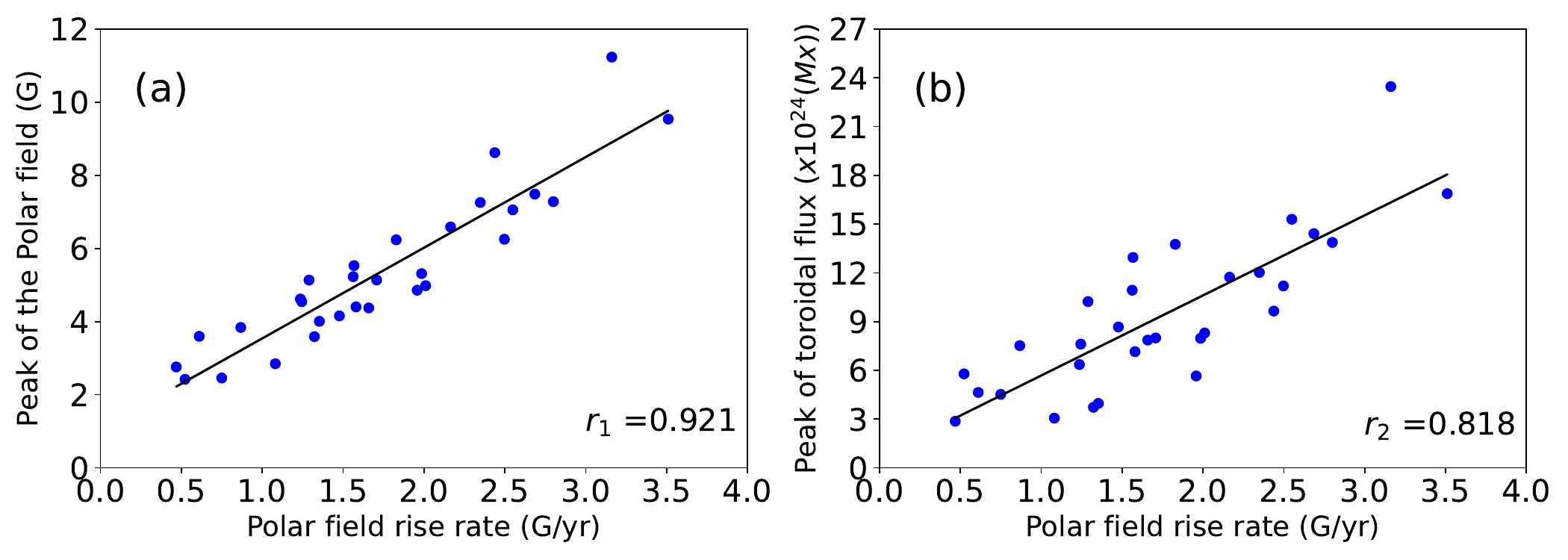}
\caption{ The scatter plots of the polar field rise rate with (a) the peak of the polar field, and (b) the peak strength of the toroidal field to be produced in the following cycle.
}
\label{fig:corr}
\end{figure}

Considering the above-mentioned factors, we perform our simulations for the following cases to test our method of solar cycle prediction against different stochastic processes present throughout different phases of the cycles:
\begin{itemize}
    \item[I)] Variation in cycle amplitudes with BMR tilts strictly following Joy's law.
    \item[II)] Same as Case I, but with varying meridional flow speed in each cycle.
    \item[III)] Variation in the BMR tilt scatter prperties with anti-Hale BMRs present in all the phases of the cycles.
    \item[IV)] Same as Case III, but with all anti-Hale BMRs in the rising phases of the cycles.    
    \item[V)] Same as Case III, but with all anti-Hale BMRs in the declining phases of the cycles.  
    \item[VI)] Same as Case III, but with the BMR tilt scatter dependent on the area of the BMRs.
\end{itemize}

We perform simulations for 30 cycles in each of these cases. The amount of the anti-Hale, and anti-Joy (present throughout all the phases of the cycles) BMRs has been taken to be within around 3\%-7\% and 10\%-30\% respectively for the last four cases.

For the limitation of space, here we present results for only the case-III and we mention the correlation coefficients for the rest of the cases in the \Tab{tab:tab1}.

\begin{wraptable}{r}{4cm}
    \begin{tabular}{ccc}
    \hline
         Cases&  $r_1$& $r_2$\\
    \hline
         I)&  0.933& 0.883\\
         \hline
         II)&  0.929& 0.867\\
         \hline
         III)&  0.921& 0.818\\
         \hline
         IV)&  0.924& 0.830\\
         \hline
         V)&  0.837& 0.774\\
         \hline
         VI)&  0.903& 0.851\\
         \hline
         
    \end{tabular}
    \caption{The correlation coefficients similar to the \Fig{fig:corr} for all the six cases mentioned.}
    \label{tab:tab1}
\end{wraptable}

Here, in the \Fig{fig:pol}, the results from a typical simulation of a cycle from case III is presented. Panel (a) shows the evolution of the surface radial field from the diffused BMRs that get transported towards the polar regions under the action of meridional circulation. Pane (b) shows the polar field strength calculated from the average field strength of the polar cap regions ($55^\circ<|\lambda|<90^\circ$). The simulated results reproduce the observed polar field evolution reasonably well. The black and the green vertical lines in panel (b) show the times of the polar field reversal ($T_r$) and the time of prediction ($T_p = T_r+ 3$) in our method. We calculate the average rise rate of the polar field build-up within the time range between $T_r$ and $T_p$ and analyse its relation with the peak of the polar field and the strength of the toroidal field in the next cycle as computed from \eqref{tor} for the above mentioned six cases. It can be easily seen that the time of prediction $T_p$ is a few years earlier than the cycle minimum, which significantly extends the time window of cycle prediction.    

The \Fig{fig:corr} shows the scatter plots of the polar field rise rate with the (a) peak of the polar field and the (b) peak of the toroidal flux for the cycles of case III. The significantly high values of correlation coefficients ($r_1$ and $r_2$) between these quantities indicate that a reliable prediction of the strength of the solar cycle can be possible from the polar field rise rate even in the presence of significant amount of stochastic fluctuations throughout the cycles.

Here in \Tab{tab:tab1} the correlation coefficients for all the six cases are mentioned. It is clear that the correlation stays high for all the cases, except case V, where all the anti-Hale BMRs are deposited in the declining phases of the cycles and that significantly disturbs the polar field build-up and eventually it hampers the correlation between these quantities. However, even in this case, the correlation is reasonable enough to get an indication of the strength of the upcoming solar cycle.

\section{Conclusion}
The results of this work suggest that the observed strong correlation of the polar field rise rate with the peak of the polar field and the peak strength of the upcoming cycle is a robust feature of the solar cycles. The extensive simulations performed in this work showed that the correlation stays high even under significant impact of stochastic fluctuations of the cycle parameters, meridional flow strength and BMR tilt properties. Hence, this feature can be utilized for an early prediction of the solar cycle much before the cycle minimum. 
Based on this method, the predicted amplitude of the Cycle 25 is $137\pm23$ \citep{KBK22}.

\section{Acknowledgements}
I thank Dr. Robert Cameron for kindly providing the SFT code and for fruitful discussions regarding the work.
I also thank the International Astronomical Union for the award of the generous travel grant to attend the in-person IAU Symposium 365, in Yerevan Armenia. The financial support from the University Grants Commission, Govt. of India is gratefully acknowledged.

\bibliographystyle{iaulike}
\bibliography{iauposter} 
\end{document}